# Structural behaviour of BiFeO$_3$/SrRuO$_3$ superlattices: an X-ray diffraction and Raman spectroscopy investigation

S. Yousfi, M. El Marssi, H. Bouyanfif*

Laboratoire de Physique de la Matière Condensée, Université de Picardie Jules Verne, 33 Rue Saint Leu, 80039 Amiens, France

**Abstract**

Epitaxial BiFeO$_3$/SrRuO$_3$ superlattices have been grown by pulsed laser deposition on a (001) oriented LaAlO$_3$ substrate and probed by X-ray diffraction and Raman spectroscopy. To investigate the structural competition between rhombohedral BiFeO$_3$ and orthorhombic SrRuO$_3$ the total thickness of all SLs was kept constant and the bilayer thickness (period) $\Lambda$ was varied. The interlayer strain effects are therefore tuned from large strain effects (short $\Lambda$ period) to quasi-relaxed structure (large $\Lambda$). A complementary investigation using X-ray diffraction and phonon dynamics hints to change from a rhombohedral to a tetragonal structure in the superlattices with the increase of the interlayer strain effect.

**Introduction**

BiFeO$_3$ (BFO) is under intense investigation because of the room temperature multiferroic state and the recent discovery of an anomalous photovoltaic response opening the way to many applications in information storage, optoelectronics and energy harvesting technologies [1, 2]. Bulk BFO belongs to the R3c space group at room temperature and can adopt under thin film form a fantastic variety of phases [3,4]. The Curie temperature and the simultaneous Pnma to R3c transition is observed at 1100K while the Neel temperature is 640K. BFO adopts in bulk a pseudo-cubic unit cell of 3.96 Å at room temperature and a robust spontaneous polarization along the [111] direction (Ps about 100$\mu$C/cm$^2$). Thin films have shown a massive potential to engineer, manipulate and create new functionalities in ferroic complex oxides through strain and low-dimensionality effects and there is an increasing focus on superlattices (SLs) for the design of new materials with remarkable properties. SLs are also ideal platforms to investigate the competition between different interactions, which is the main objective of this work. We chose to combine the antiferromagnetic-ferroelectric BFO with the ferromagnetic-paraelectric SrRuO$_3$ (SRO) in such SLs and focus this report on the structural interaction at the heterointerfaces of the multilayers. A similar Metal/Ferroelectric SrRuO$_3$/PbTiO$_3$ type of superlattice has been recently investigated and an exotic hierarchical domain structure (so called super-crystal) has been revealed [5]. PbTiO$_3$ being a classical ferroelectric it would be interesting to explore similar super crystal structure but with a multiferroic such as BFO. Being motivated by the fascinating work by Pavel's group [5] this preliminary structural investigation aims on a long term perspective to explore the possibility to obtain similar super-crystal state in multiferroic BFO/SRO SLs. Similar to PbTiO$_3$/SrRuO$_3$ SLs the possibility to electrically and magnetically tune the electronic and transport properties of such super-crystal multiferroic state in BFO/SRO SLs is envisioned. Topological spin and polar textures (skyrmions and vortex) with robust electric/magnetic properties may also be expected in such confined nanoscale structures. While BFO is a rhombohedral R3c system, SRO shows an orthorhombic structure with a Pnma space group. The two oxygen superlattices also differ with a$^-$a$^-$a$^-$ and a$^-$a$^-$c$^+$ oxygen rotation/tilt system for respectively BFO and SRO. Symmetry mismatch of the oxygen framework at the heterointerfaces is therefore also expected. SRO is a metallic system often used as an electrode and shows a ferromagnetic order below 160K [6][7]. The pseudo-cubic lattice parameter of SRO is 3.93Å in bulk but may be modified in thin film due to the strain imposed by the substrate [8]. In this report, eight SLs are investigated with a total thickness of about 5000Å. The strategy to tune the interlayer interaction consists of keeping the same ratio of BFO and SRO in the bilayer and varies the bilayer thicknesss (Λ) while maintaining the total thickness constant. To do so SLs with different numbers of bilayers are therefore grown.

**Experiment**

The SLs were grown by pulsed laser deposition (excimer KrF : 248nm) at 6Hz pulse frequency, a fluence of 1.8J/cm$^2$, 725°C and oxygen pressure of 5.10$^{-2}$mbar. We used a BFO target enriched with 10% Bismuth and a Fe substitution with 5% of manganese (Bi$_{1.1}$Fe$_{0.95}$Mn$_{0.05}$O$_3$). This BFO target stoichiometry is chosen to prevent Bi loss during the growth and to decrease the amount of leakage currents. The bottom electrode SRO (20nm) was synthesized at 650°C and oxygen pressure 0.3mbar. The crystalline and epitaxial structure were studied by high resolution X-ray diffraction (Bruker HRXRD d8 and λ=1.54056Å). Raman spectroscopy was performed using a Renishaw spectrometer with a laser excitation at 514nm, in backscattering geometry, with two polarisations $Z(YY)\bar{Z}$ and $Z(XY)\bar{Z}$. (X,Y,Z) are the laboratory axes with Z perpendicular to the substrate surface.

**Results and discussions**

Table 1 below presents the characteristics of the different SLs. All the SLs start and end with a BFO layer to get a symmetric stacking. From the rate deposition of the single layers, we chose to fix the BFO ratio within each period Λ to about Λ/3 and SRO ratio to 2Λ/3. The 8 SLs are deposited on $LaAlO_3$ (001) (LAO) substrate buffered with 20nm thick SRO bottom electrode.

| Number of bilayers | Period Λ (Å) |
|---|---|
| 6 | 831 |
| 8 | 645 |
| 10 | 515 |
| 12 | 424 |
| 15 | 281 |
| 20 | 269 |
| 30 | 178 |
| 60 | 75 |

Table 1. Characteristics of the superlattices.

Figure 1 presents the X-ray diffraction pattern in θ/2θ geometry for the 8 SLs on a very large range of diffraction angles. No parasitic phases are observed within the resolution limit. Intense, sharp and regularly spaced satellite peaks are observed which demonstrate a true chemical modulation of the stacking and not a solid solution. As expected, the spacing of the satellite peaks decreases when the period Λ increases. Total thickness and thickness of the bilayers can be deduced from the relation:

$$\Lambda = \frac{\lambda}{2(\sin\theta_{n+1} - \sin\theta_n)} \quad (1)$$

where $\theta_{n+1}$ and $\theta_n$ are two successive satellite peaks and λ is the X-ray wavelength (λ=1.54056Å). Results are given in table 1 above. Figure 2 shows ω scan (rocking-curve) on the most intense satellite peak. For all SLs, the Full Width at Half Maximum (FWHM) are between 0.3° and 0.4° indicating a good crystalline orientation of all SLs (0.06° for the LAO substrate).

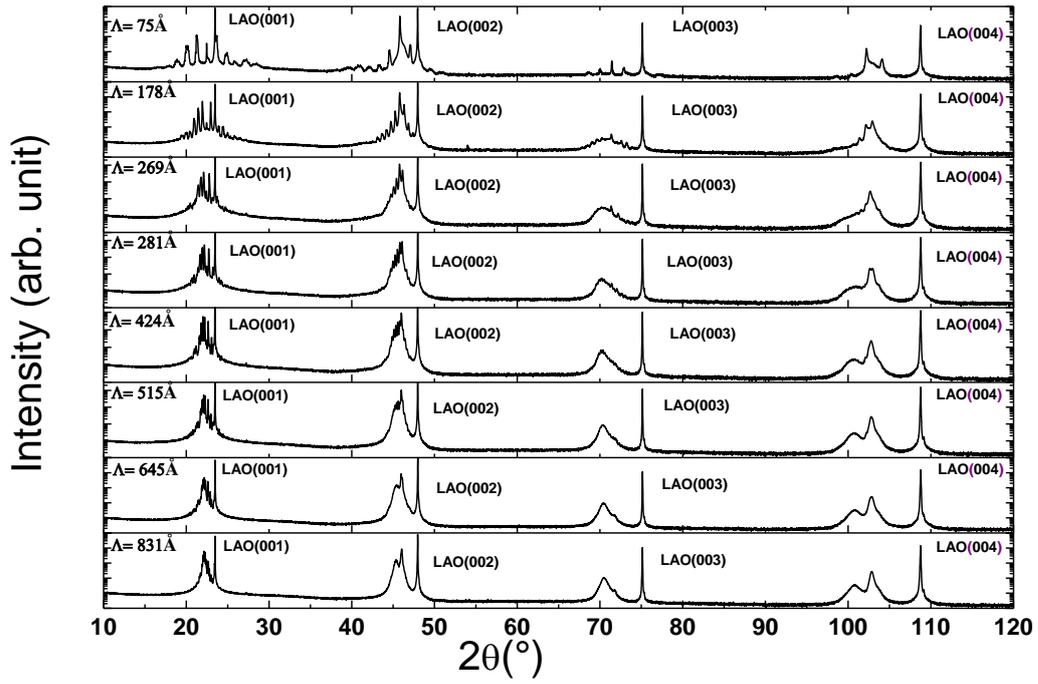

Figure 1. (a) The θ/2θ diffraction pattern for the SLs with different periods Λ. The periodicity Λ is deduced from relation (1) and the angular distance between two consecutive satellite peaks (see table 1).

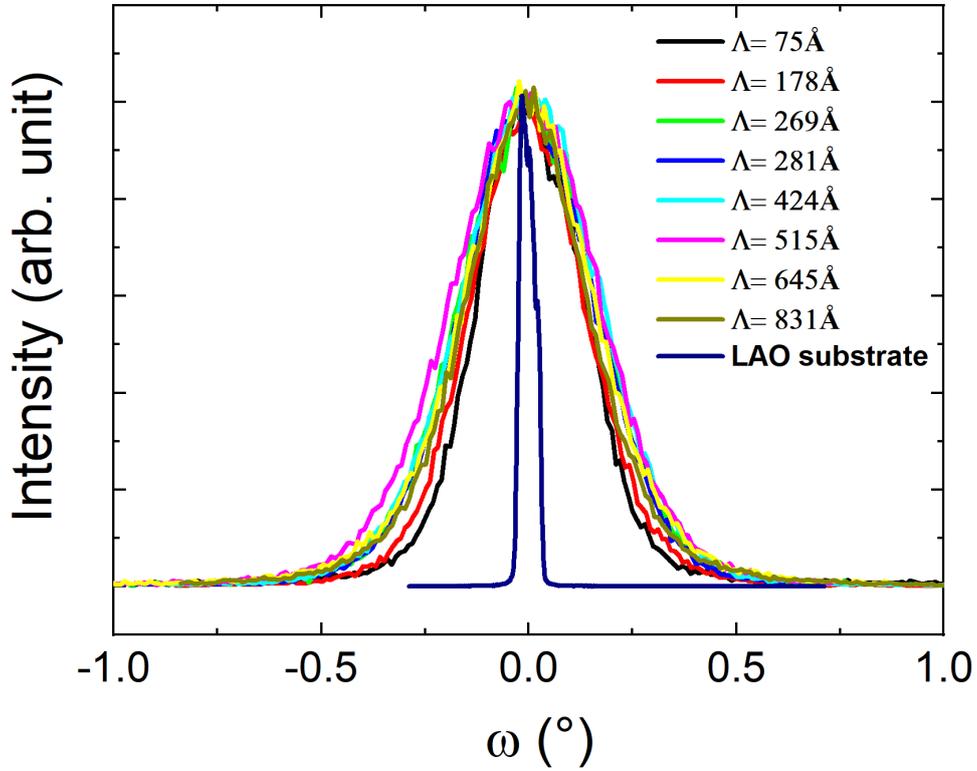

Figure 2. Intensity normalized rocking curves obtained on the most intense satellite peak ("average" second order next to the LAO(002) reflection) of SLs.

In order to get access to the out of plane lattice parameters of the BFO and SRO layers and the exact number of BFO and SRO unit cells, these diffractograms have been simulated with the Matlab InteractiveXRDfit program developed by Dr. C. Lichtensteiger [9]. This semi-kinematical approach allows a fast simulation by adjusting the number of unit cells and the out of plane lattice parameters in each layer (BFO and SRO) of the period Λ. The comparison between the simulations and the experiments is shown on Figure 3 for the 1st and 2nd order of diffraction of LAO and for three representatives selected superlattices (6 bilayers, 15 bilayers and 60 bilayers).

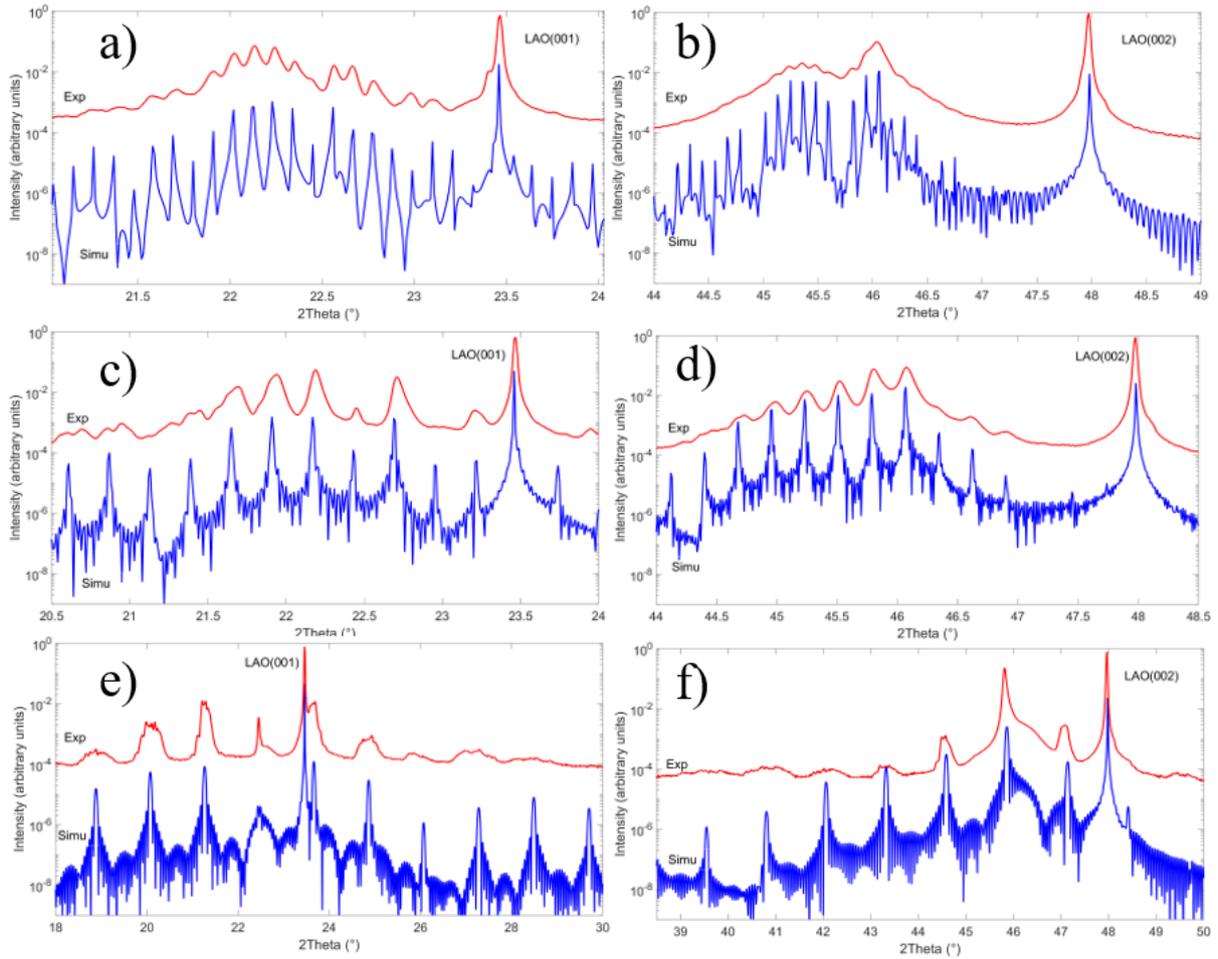

Figure 3: Comparison between experiments (Exp) and simulation (Simu) of diffractograms corresponding to 6 bilayers ($\Lambda$=831 Å and top a and b), 15 bilayers ($\Lambda$=281 Å middle c and d), 60 bilayers ($\Lambda$=75 Å and bottom e and f) around first and second order of diffraction of the LaAlO$_3$ substrate (LAO(001) and LAO(002)).

The overall correspondence is good considering the possible disorders existing in non-ideal superlattices (limited crystalline coherence, interdiffusion, roughness for instance) that are not taken into account in the simulation (peaks are extremely sharp compared to the experimental ones). Nevertheless, such simulations give access to important information such as the out of plane lattice parameters and the number of unit cells along the growth direction (thickness of BFO and SRO layers). The extracted structural information is given on table 2 below for the whole set of SLs. The value of $\Lambda$ is close to the one expected from the rate deposition and the number of laser shots on each target. $\Lambda$ is experimentally evaluated from such simulations and values in table 1 are more reliable than the expected ones.

| Number of bilayers | C_SRO (Å) | Number of cell SRO | C_BFO (Å) | Number of cell BFO | Period Λ (Å) | Total Thickness (Å) |
|---|---|---|---|---|---|---|
| 6 period | 3.940 | 150 | 4.000 | 60 | 831 | 4986 |
| 8 period | 3.940 | 115 | 4.005 | 48 | 645 | 5160 |
| 10 period | 3.940 | 90 | 4.000 | 40 | 515 | 5150 |
| 12 period | 3.940 | 72 | 4.000 | 35 | 424 | 5088 |
| 15 period | 3.940 | 46 | 4.010 | 25 | 281 | 4215 |
| 20 period | 3.942 | 50 | 4.014 | 18 | 269 | 5380 |
| 30 period | 3.937 | 36 | 4.050 | 9 | 178 | 5340 |
| 60 period | 3.920 | 14 | 4.050 | 5 | 75 | 4500 |

Table 2: Structural information deduced from the simulations of the diffractograms.

We observe on table 2 an increase of the out of plane lattice parameter of the BFO layers on decreasing the period Λ while the out of plane of SRO shows a decrease for very small Λ. The structural behaviour is discussed more in detail below.

X-ray reflectivity is a technique sensitive to the electronic density (chemical composition), thickness and roughness and is a straightforward method to quickly evaluate the thickness of films and multilayers by using the following relation:

$$\theta^2 - \theta_c^2 = m^2 \left(\frac{\lambda}{2\Lambda}\right)^2 \quad (2)$$

With Λ the bilayer thickness, λ the X-ray wavelength, θ the satellite maxima, m the order of reflection and $\theta_c$ the critical angle of reflectivity. By plotting the maxima θ² versus the order m² and from a linear fitting, a rapid evaluation of the bilayer Λ can be made from the slope of the linear fit. Note that the distance between the reflectivity maxima depends only on the thicknesses and not on the roughness. This simple method allows therefore to determine the thicknesses without simulating the whole intensity profile. Figure 4 displays X-ray reflectivity scans obtained on our SLs. Only satellite maxima from bilayer thicknesses are observed on Figure 4 and no maxima (Kiessig fringes in between the satellites) from the SLs total thickness is evidenced within the resolution limit of our diffractometer.

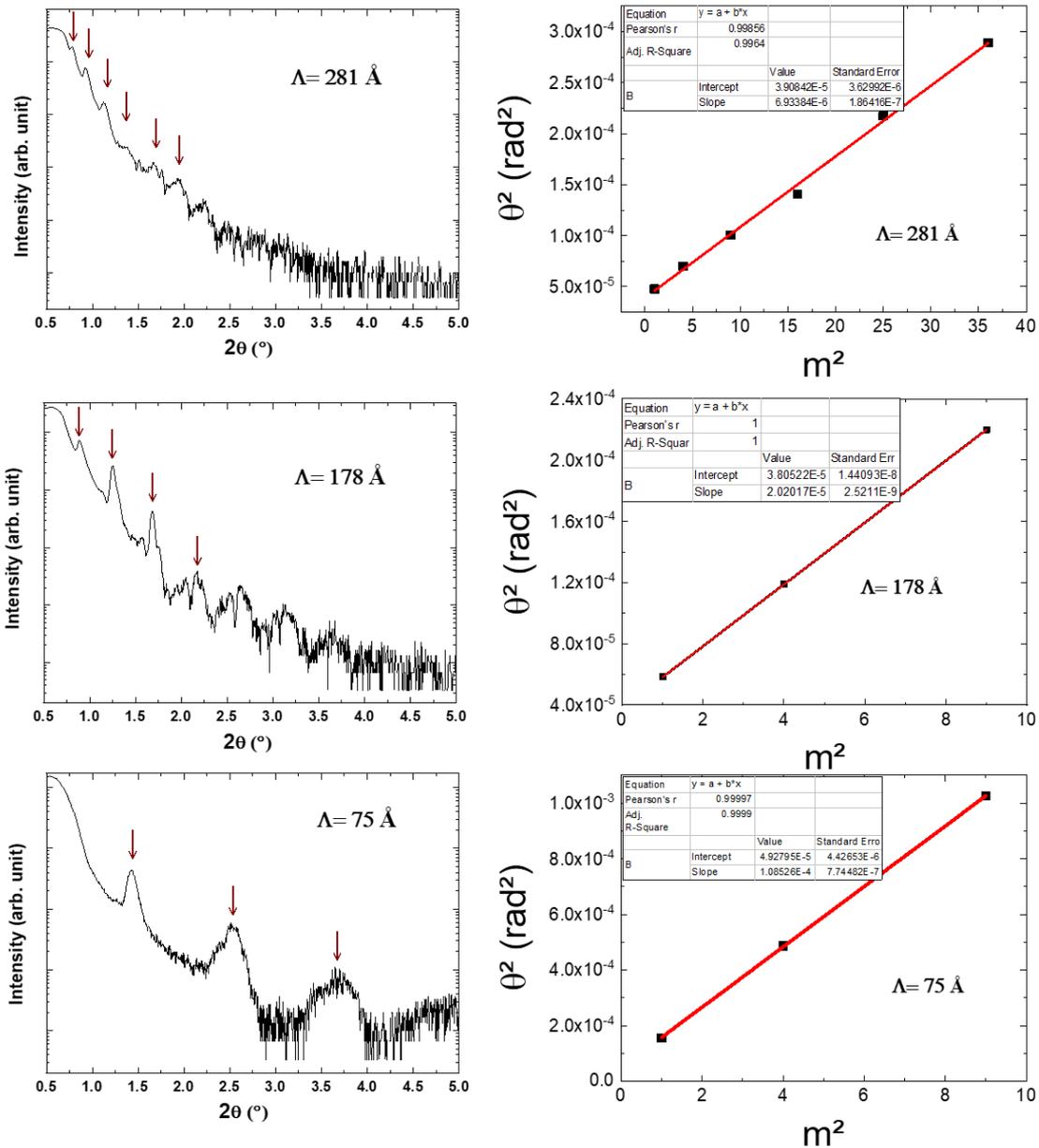

Figure 4: X-ray reflectivity scan on three representative SLs from 15 to 60 Bilayers (left row). Right row: Linear fit of the maxima position θ² as a function of m² for the SLs (m:order of diffraction, no dimension). Symbols correspond to experimental data while the continuous line is the fit (y=a+b*x with b:slope and a:intercept) and the inserted table gives fitting parameters.

Satellite maxima are well pronounced at low angles while no intensity modulations are detected at high angles. Roughness and low resolution of the diffractometer explain the high angle disappearance of reflectivity oscillations. The lower the number of bilayers the larger the number of satellites. The angular separation is inversely proportional to the bilayer thickness and an estimation of such bilayer thickness is therefore possible. Figure 4 also presents the linear fit based on relation (2) for three SLs. The slope of the linear fit given in figure 4 increases with the number of bilayers (recall that total thickness of SLs is kept constant). Equivalently, the slope estimated on figure 4 is inversely proportional to the thickness bilayer. The values are $6.93.10^{-6}$, $2.02.10^{-5}$ and $1.09.10^{-4}$ respectively for 15, 30 and 60 bilayers. It allows us to estimate the bilayer thickness ($\Lambda$=292 Å for 15 bilayers ; $\Lambda$=171Å for 30 bilayers; $\Lambda$=74 Å for 60 bilayers). The values are in good agreement with $\Lambda$ deduced from the simulation of the θ-2θ diffraction patterns and given in table 2 despite the very complex profile of the reflectivity scans that we explain by the roughness (top surface and at interfaces), interdiffusion and the strong mosaicity of the $LaAlO_3$ substrates (ferroelastic twins misoriented). The reflectivity and diffraction measurements allow us to determine and confirm the period $\Lambda$ and total thickness values of the SLs.

To better investigate the structural evolution of the SLs we used X-ray reciprocal space mapping (RSM) around the (103) and (113) family of planes. Symmetry and domain structures dictate the relative position and number of reflections for such family of planes and we recall on Fig. 5 the expected results for the different symmetries that BFO can present depending on the in plane compressive strain state. Number of domains and reflections decrease on increasing symmetry.

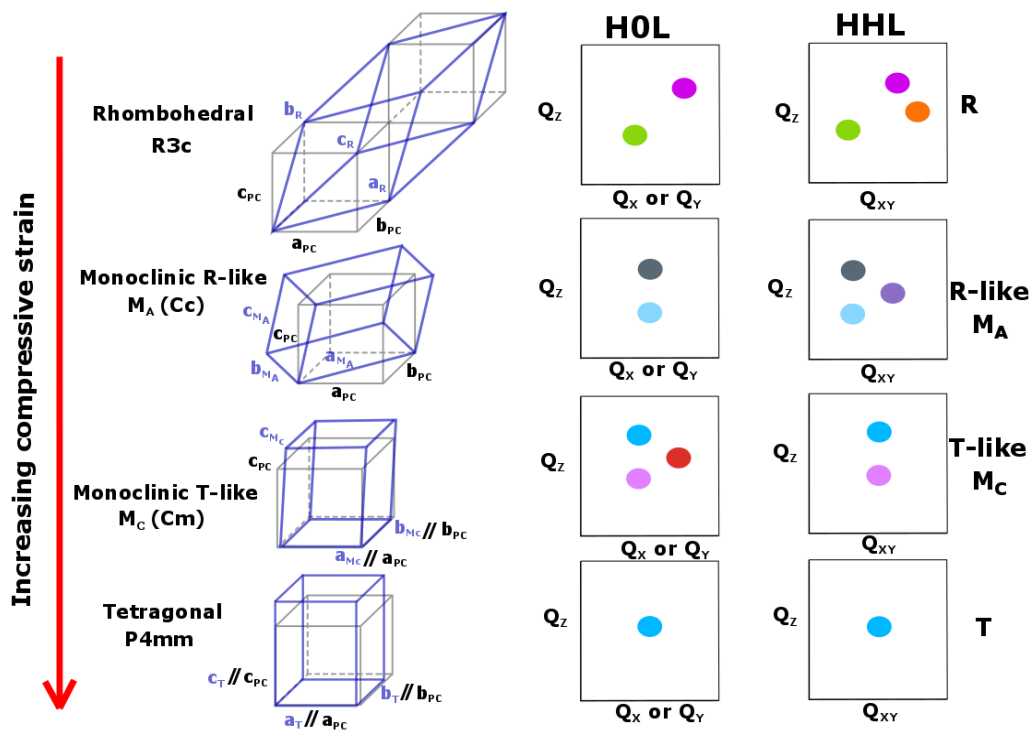

Figure 5: Relation between the symmetry, the number and relative position of the reflection for reciprocal space mapping around the (H0L) and (HHL) family of planes. Small compressive

strain is observed for thick and relaxed BFO layers while strong in plane compressive strain are observed for thin and unrelaxed BFO layers (when grown on substrates with suitable lattice parameters and compressive epitaxial mismatch). Note that Rhombohedral thin films often show only two peaks in (HHL) neither aligned in $Q_{xy}$ nor in $Q_z$ [4,10] due to preferential growth of certain types of domains.

Figure 6 shows the RSM results obtained for five SLs. Considering that the whole SL structure diffracts, it is not possible to deduce the structure/symmetry of the BFO and SRO layers separately and only a global symmetry can be inferred from such RSM. The $Q_x$ and $Q_{xy}$ positions for the SLs and the substrate (not shown for clarity) are different implying a different in plane lattice parameters and a partially relaxed strain induced by the substrate. The two sets of (103) and (113) RSM show broad reflexion synonymous of small domains and mosaicity. The RSM show a certain evolution on decreasing the periodicity $\Lambda$. The RSM (103) and (113) for large periods ($\Lambda$=831Å and $\Lambda$=645Å) present two reflexions (located with symbols +) not aligned in $Q_z$, $Q_x$ and $Q_{xy}$. Doublet in RSM neither aligned in $Q_z$ nor in $Q_{x(xy)}$ strongly suggest a rhombohedral symmetry for the SLs (see Fig. 5 and Ref.10). In contrast the (103) and (113) RSM corresponding to short periodicity $\Lambda$ are showing regularly spaced satellites aligned in $Q_x$ and $Q_{xy}$, without any splitting of the nodes along in plane directions. This strongly implies a tetragonal deformation for the SLs with short periodicity. This evolution from a rhombohedral to a tetragonal structure is induced by the increasing interlayer strain effects when the periodicity $\Lambda$ decreases. Rhombohedral to tetragonal change has already been observed in BFO film but never to our knowledge in SLs [10]. Note that SRO thin films tend to adopt a tetragonal symmetry under compressive in plane strain [11]. It is not clear whether super-tetragonal distortion exists in the BFO layers but we notice a non-zero structure factor and therefore satellite peaks centred at about $\theta$=39° for the SL with ultra-short $\Lambda$=75Å in Fig. 1. This position coincides with the position of the BFO diffraction peak in the super-tetragonal phase and only a TEM analysis can confirm this hypothesis [10].

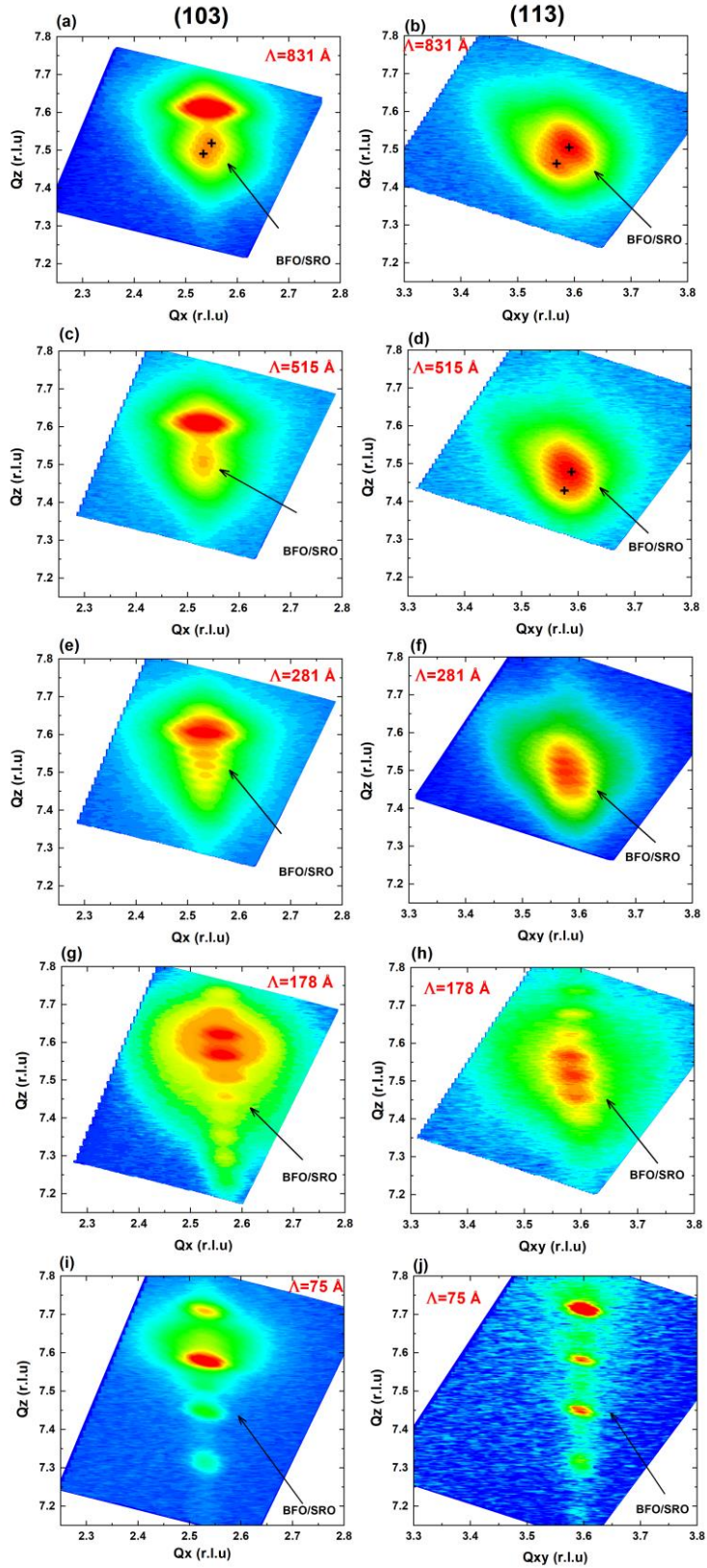

Figure 6. Reciprocal space mapping around (103) and (113) for SLs with Λ=831 Å (a,b) Λ=515 Å (c,d) Λ=281 Å (e,f) Λ=178 Å (g,h) and Λ=75 Å (i,j).

To fully exploit the X-ray diffraction experiments we plot the out of plane lattice parameters (see table 2) and the in plane lattice parameters deduced from the RSM on figure 7 (a). While the in plane lattice parameters are constant whatever the period Λ, an increase (decrease) of the out of plane lattice parameter of BFO (SRO) is observed when the period Λ decreases. This strongly suggests an increase of the compressive in plane strain for BFO. Opposite trend is therefore deduced for SRO which is under in plane tensile strain. Such results are naturally expected from the bulk pseudo cubic lattice parameter values of BFO and SRO (3.96Å for BFO and 3.93Å). These bulk pseudo cubic lattice parameters imply a tensile in plane strain imposed by BFO on SRO or equivalently, a compressive in plane strain imposed by SRO on BFO. Concomitant to this increasing out of plane lattice parameter for BFO a tetragonal distortion is indeed evidenced. This increase of the tetragonal distortion is confirmed by the ratio $c_{pc}/a_{pc}$ evolution with the period Λ on figure 7 (b). The BFO ratio $c_{pc}/a_{pc}$ is close to 1.015 for large period and increases up to 1.035 for short period. The SRO ratio is, however, close to one for the whole range of Λ except for Λ=75 Å and a value below one. A region with a rhombohedral structure transforms therefore into a region with a tetragonal structure with in between a continuous change for BFO. This smooth change of structure is inferred from the continuous change of the structural lattice parameters reminiscent of morphotropic phase boundaries. It is important to bring to the attention of the reader the observation of morphotropic like phase boundaries in superlattices made of $PbTiO_3/CaTiO_3$ and similar structural behaviour may take place in the $BiFeO_3/SrRuO_3$ SLs [12]. Transmission electron microscopy is under investigation to further confirm the observed structural changes.

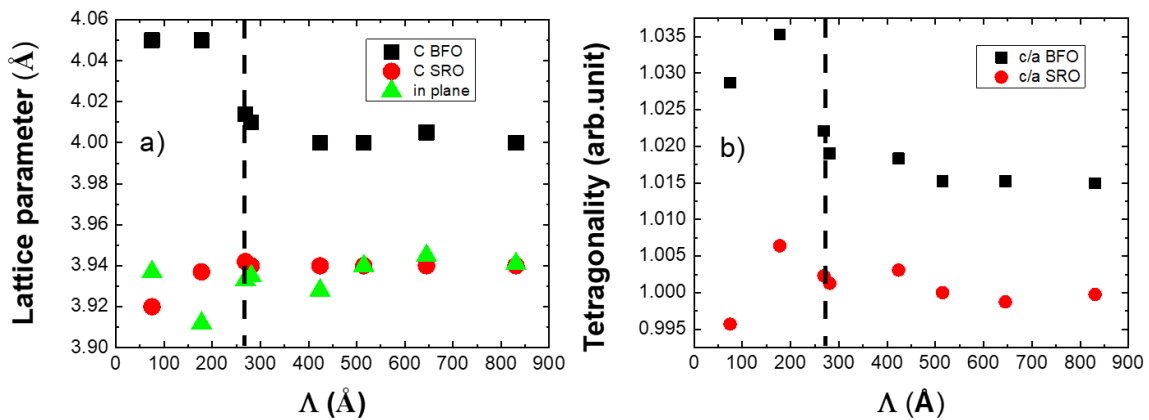

Figure 7 (a) Out of plane and in plane lattice parameter evolution with the period Λ (b) The evolution of the ratio $c_{pc}/a_{pc}$ with the period Λ. The vertical dashed lines separate the region of tetragonal like state (larger $c_{pc}/a_{pc}$) from the rhombohedral like state. Error bars are within the marks.

To confirm the structural changes detected via X-ray diffraction the symmetry of the SLs was further scrutinized through the prism of lattice dynamics using Raman spectroscopy. The Raman spectra of the SLs collected in backscattering geometry with two geometries of measurements $Z(YY)\bar{Z}$ (parallel) and $Z(XY)\bar{Z}$ (crossed) are presented in Fig. 8. An evident influence of the period Λ is observed on the Raman spectra of the SLs. The Raman spectra for the SL with short period (Λ=75Å) being completely different from the Raman spectra of the SL with large period (Λ=831Å) in both crossed and parallel geometries. This difference is a clear signature of a Λ driven symmetry change in these epitaxial superlattices. For Λ=831Å most intense phonons mode are detected at 141cm$^{-1}$, 174cm$^{-1}$ and 222cm$^{-1}$ in $Z(YY)\bar{Z}$ geometry. These phonon frequencies are those observed for A$_1$ modes in BFO rhombohedral symmetry is in agreement with XRD analysis and an identification of a rhombohedral symmetry based on the RSM [13, 14]. These three A$_1$ phonons (141cm$^{-1}$, 174cm$^{-1}$ and 222cm$^{-1}$) progressively disappear on decreasing Λ. The Raman spectra thus suggest a R3C symmetry in the BFO layers that disappears for ultra-short periods. Phonons are also detected in the range 350 cm$^{-1}$- 400cm$^{-1}$ and are characteristic of the vibrational Raman spectra of SRO [17, 18]. It is, however, difficult to make conclusion on the SRO structural behaviour from the Raman spectra.

Phonons observed in parallel geometry are different to those in crossed geometry. Such polarized Raman spectra are typical of epitaxial layers and superlattices and enable the discussion of Raman selection rules. Raman intensity in crossed geometry is weak and only one phonon is visible at 279cm$^{-1}$ for Λ=831Å. This phonon is attributed to an E mode of the R3c symmetry as observed in BFO thin films and crystals [13,14]. On increasing the period this E mode disappears. The disappearance of the R3c phonon excitation is a signature of the symmetry change in the SLs in agreement with the XRD analysis. Excitations above 600cm$^{-1}$ are also observed in the SLs and deserve some attention. An intense phonon mode is indeed evidenced at 622cm$^{-1}$. The origin of this phonon is not yet clear and may be connected to a Jahn-Taller distortion of the oxygen octahedral induced by the 5% iron (Fe) substitution by the manganese (Mn) [15,16]. The intensity for this mode decreases for SLs with shorter period and we observe the simultaneous appearance of another vibrational mode at 720cm$^{-1}$. Such spectral change clearly indicates a local structural change of the oxygen octahedral environment in concordance with the rhombohedral to tetragonal like change evidenced above by XRD. We note the quasi-extinction for all the vibrational modes for the SLs with shorter period. The Raman selection rules analysis of the tetragonal symmetry P4mm, the monoclinic symmetry Cc and the R3c symmetry is provided in table 3 [19]. Raman selection rules indicate a decrease of the number of phonons when going from the R3c to the Cc and P4mm symmetries. In particular no phonons are expected for the tetragonal P4mm symmetry in crossed geometry similarly to our observation on figure 8 (b) for the ultra-short period. The Cc symmetry, already observed in BFO strained thin film, is also correlated to weak phonon bands in crossed geometry and it is not clear which symmetry (Cc versus P4mm) is present in our SLs [4]. Nevertheless a clear disappearance of the R3c in profit to a tetragonal-like distortion is evidenced in both Raman spectra and X-ray diffraction in SLs with ultra-short periods.

| Mode | $(xx)_c$ $(xx)_t$ $(x'x')_m$ | $(yy)_c$ $(yy)_t$ $(y'y')_m$ | $(xy)_c$ $(xy)_t$ $(x'y')_m$ | $(x'x')_c$ $(x'x')_t$ $(xx)_m$ | $(y'y')_c$ $(y'y')_t$ $(yy)_m$ | $(x'y')_c$ $(x'y')_t$ $(xy)_m$ |
|---|---|---|---|---|---|---|
| $A_1$(R3c) | $\frac{1}{9}(2a+b)^2$ | $\frac{1}{9}(2a+b)^2$ | $\frac{1}{9}(a-b)^2 \approx 0$ | $\frac{1}{2}\left[a^2 + \frac{1}{9}(a+2b)^2\right]$ | $\frac{1}{2}\left[a^2 + \frac{1}{9}(a+2b)^2\right]$ | 0 |
| E(R3c) | $\frac{4}{9}(-c+\sqrt{2}d)^2$ | $\frac{4}{9}(-c+\sqrt{2}d)^2$ | $\frac{1}{9}(2c+\sqrt{2}d)^2$ | $\frac{1}{2}\left[c^2 + \frac{1}{9}(c+2\sqrt{2}d)^2\right]$ | $\frac{1}{2}\left[c^2 + \frac{1}{9}(c+2\sqrt{2}d)^2\right]$ | $\frac{1}{3}(-c+\sqrt{2}d)^2$ |
| $A_1$(4mm) | $a^2$ | $a^2$ | 0 | $a^2$ | $a^2$ | 0 |
| $B_1$(4mm) | $c^2$ | $c^2$ | 0 | 0 | 0 | $c^2$ |
| A'(Cc) | $\frac{1}{4}(a+b)^2$ | $\frac{1}{4}(a+b)^2$ | $\frac{1}{4}(a-b)^2 \approx 0$ | $\frac{1}{2}(a^2+b^2)$ | $\frac{1}{2}(a^2+b^2)$ | 0 |
| A''(Cc) | $e^2$ | $e^2$ | 0 | 0 | 0 | $e^2$ |

Table 3. The Raman selection rules analysis for rhombohedral (R3c), tetragonal (P4mm) and monoclinic (Cc) symmetry [19].

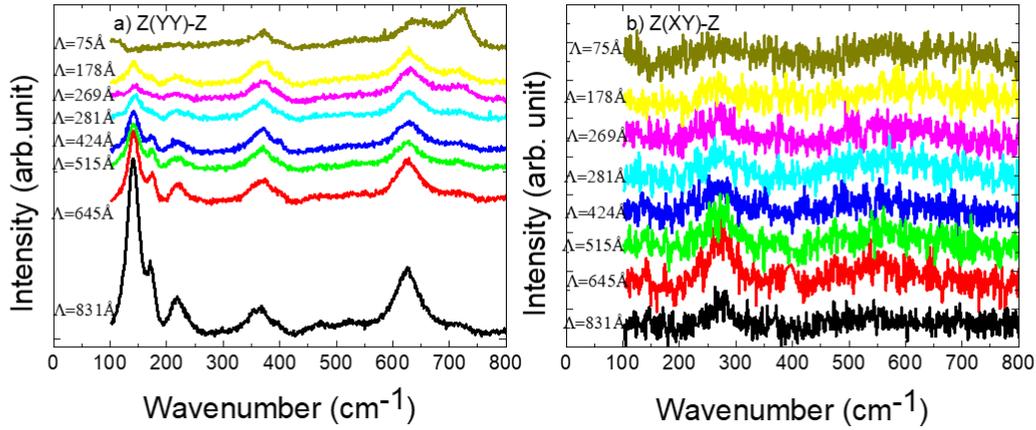

Figure 8. Raman spectra of the SLS in (a) parallel $Z(YY)\bar{Z}$ geometry and (b) crossed $Z(XY)\bar{Z}$ geometry. Note the large difference in intensity between parallel and crossed geometries (crossed spectra seem noisy despite many attempts to improve the signal over noise ratio and we infer therefore that it is a direct impact of the underlying symmetry of the SLs).

Finally we note that the observed phonons are separately classified into BFO and SRO layers, which points into the non-propagating nature of the observed excitations. The detected optic phonons are therefore confined within the layers and this is caused by non-superposition of the phonon branches in the Brillouin zone of the BFO and SRO constituents.

**Conclusion**

Superlattices were used in this report to investigate the structural competition at heterointerfaces between BFO and SRO materials. In order to tune the interaction, epitaxial superlattices with different bilayer thickness but constant total thickness was grown by ablation laser and probed by X-ray diffraction and Raman spectroscopy. Increasing the strength of the structural interaction results into a change of the global symmetry from a Rhombohedral like into a Tetragonal like state. Confined optic phonons reminiscent of BFO and SRO were detected and Raman selection rules confirm the observed symmetry changes. Epitaxial mismatch at the heterointerfaces is at the heart of this behaviour and transmission electron microscopy is under way to better understand the structural changes and the role of the oxygen tilt/rotation system. Magnetic measurements are also envisaged to reveal the antiferromagnetic (BFO) versus ferromagnetic (SRO) competition at the interfaces of these superlattices. Indeed, although the interlayer strain is believed to dominate the observed structural changes, magnetic characterizations are needed to evaluate the relative role of competing magnetic (AF versus F) interactions on the rhombohedral-to-tetragonal structural change.

**Acknowledgements**

The authors gratefully acknowledge the generous financial support of region of Picardy (project ZOOM) and the European Union Horizon 2020 Research and Innovation actions MSCA-RISE-ENGIMA (No. 778072) and MSCA-RISE-MELON (No. 872631).